# Front Velocity and Front Location of Lock-exchange Gravity Currents Descending a Slope in a Linearly Stratified Environment

Liang Zhao[1], Zhiguo He, M.ASCE [2], Yafei Lv[3], Ying-Tien Lin[4], Peng Hu[5], Thomas Pähtz[6]

[1]Ph.D. Candidate, Ocean College, Zhejiang University, Zhoushan 316021, China. E-mail: liangz@zju.edu.cn

[2]Professor, Ocean College, Zhejiang University, Zhoushan 316021, China; and, State Key Laboratory of Satellite Ocean Environment Dynamics, The Second Institute of Oceanography, State Oceanic Administration, Hangzhou 310012, China (Corresponding author). E-mail: hezhiguo@zju.edu.cn

[3]Graduate Student, Ocean College, Zhejiang University, Zhoushan 316021, China. E-mail: lvyafei@zju.edu.cn

[4] Associate Professor, Ocean College, Zhejiang University, Zhoushan 316021, China. Email: kevinlin@zju.edu.cn

[5] Associate Professor, Ocean College, Zhejiang University, Zhoushan 316021, China. Email: pengphu@zju.edu.cn

[6] Research Professor, Ocean College, Zhejiang University, Zhoushan 316021, China. Email: tpaehtz@gmail.com

**Abstract:** Gravity currents descending a slope in a linearly stratified environment can be frequently encountered in nature. However, few studies have quantitatively investigated the evolution process of lock-exchange gravity currents in such environments. A new set of analytical formulae is proposed by integrating both mass conservation and linear momentum equations to determine the front velocity and the front location of a downslope current. Based on the thermal theory, the formula considers the influence of ambient stratification by introducing a newly defined stratification coefficient in the acceleration stage. As for the deceleration stage, the formula is derived by adding a parameter that takes into account the density distribution of the ambient water. The transition point between the acceleration and deceleration stages and the maximum front velocity are also determined by the proposed formulae. Lock-exchange gravity current experiments are conducted in the flume with linear stratifications to provide data for the validation of the formulae. The comparisons between the calculated and measured data in terms of front location and front velocity show satisfactory agreements, which reveal that front velocity presents a rapid acceleration stage and then a deceleration stage in a linearly stratified ambient.

# Introduction

Gravity currents are flows driven by the horizontal density contrast between the current and the ambient water (Simpson 1982). They are an important phenomenon in both nature and engineering practices. Examples of gravity currents include thunderstorm outflows, salt water intrusions in estuaries, sediment-laden river discharges in lakes, and heavier pollutant discharge in water bodies, etc. (Simpson 1982; Ottolenghi et al. 2016; Steenhauer et al. 2017). Gravity currents have been extensively studied by a continuous-inflow or by a lock-exchange technique (Ho and Lin 2015; Ottolenghi et al. 2016; He et al. 2017) in a laboratory flume. A lock-exchange gravity current is usually generated by a sudden release of a locked volume of dense fluid into ambient lighter water (Ottolenghi et al. 2016) in the flume, leading to an exchange between current and ambient water. The typical structure of a lock-exchange gravity current includes a dense and semi-elliptic front head propagating in the flume, followed by a thin tail. The front location and velocity of the dense head determine the distance that the current can reach and the time the current arrives at a certain point. Therefore, quantitatively understanding the front location and front velocity is of key significance in investigating the dynamic process of lock-exchange gravity currents.

Previous studies (Huppert and Simpson 1980; Meiburg et al. 2015) have demonstrated that the propagation of a lock-exchange gravity current on a flat bed can be divided into three stages, i.e., a relatively short acceleration stage, followed by a slumping stage with a nearly constant front velocity, and then a self-similar deceleration stage. However, in reality, gravity currents often occur in varying topographic beds (Steenhauer et al. 2017), which means that the above three stages may not always exist. Several experimental studies have already shown that the current first accelerates and then decelerates when the lock-exchange gravity current propagates down a slope in a homogenous environment (Beghin et al. 1981; Maxworthy and Nokes 2007). A thermal theory (Beghin et al. 1981), assuming that the current develops from a 'virtual origin' and propagates with a constant ratio of head height to head length, was proposed to calculate front velocity and location, and this was validated by lock-exchange experiments of gravity currents moving down slopes between 5° and 90°. Furthermore, various models based on the thermal theory were developed to investigate different kinds of gravity currents, such as Non-Boussinesq gravity currents (Dai 2014; 2015), powder-snow avalanches (Rastello and Hopfinger 2004), and particle-laden currents (Dade et al. 1994). However, those models are only applicable to gravity currents in homogeneous environments.

In real geophysical environments, gravity currents also propagate in non-homogeneous/stratified environments (Baines 2001; Wells and Nadarajah 2009; Cortés et al. 2014; Snow and Sutherland 2014) since the vertical density variation of the ambient often exists, such as thermoclines in lakes and haloclines in estuaries and oceans (Fernandez and Imberger 2008; Meiburg et al. 2015). In recent years, experiments of lock-exchange gravity currents in stratified environments have also been carried out to investigate the front velocity (Snow and Sutherland 2014; He et al. 2017), the velocity field (He et al. 2017), the mixing and entrainment (Samothrakis and Cotel 2006), and the separation depth (Snow and Sutherland 2014; He et al. 2017). By fitting with experimental data, Maxworthy et al. (2002) proposed empirical formulae to determine the front velocity of gravity currents using Froude numbers in a flat and linearly stratified environment, which were also validated by comparing with the results from direct numerical simulations. Based on the steady-state theory proposed by Benjamin (1968), Ungarish (2006) discussed the Froude number of gravity currents on a flat bed in weak and strong linearly stratified environments, which were further tested and discussed by Birman et al. (2007) using numerical simulations. Most recently, Longo et al. (2016) conducted experiments of gravity currents in a flat and linearly stratified environment in both rectangular and semi-circular channels to validate a simplified theoretical model to describe the front velocity at the slumping stage. Moreover, He et al. (2017) experimentally studied the hydrodynamics of lock-exchange gravity currents down a slope in a linearly stratified environment and extended the thermal theory to examine current velocity in the deceleration stage, however, the velocity in the acceleration stage was neglected.

Therefore, although some studies have documented the hydrodynamics of a lock-exchange gravity current, an entirely quantitative study of its propagation has not yet been fully investigated, especially in an inclined and stratified environment. The objective of this study is to develop a set of formulae to determine the front velocity and front location of lock-exchange gravity currents down a slope in linearly stratified environments by considering both the ambient density variation and water entrainment. The proposed formulae are further validated by comparing with the experimental data in both relatively weak and strong ambient stratifications. This paper is organized as follows. First, we describe the experimental set-up and results. Then, we derive the formulae used to determine the development of the gravity current and validate these formulae with the experimental data. Finally, some discussions and conclusions are presented.

# Experiments and Parameters

*Experimental set-up and procedure*

**Table 1**. Experimental runs and the relevant parameters

| Run | $\theta$ | $S$ | $\rho_{c0}$ (kg/m³) | $\rho_B$ (kg/m³) | $\rho_s$ (kg/m³) | $m$ (1/m) | $g_0'$ (m/s²) | Re |
|---|---|---|---|---|---|---|---|---|
| 1 | 6 | 0.66 | 1019.57 | 1014.08 | 1003.58 | 0.0044 | 0.155 | 3515 |
| 2 | 6 | 1.31 | 1011.73 | 1013.98 | 1004.52 | 0.0039 | 0.070 | 2311 |
| 3 | 6 | 2.55 | 1009.10 | 1016.09 | 1004.60 | 0.0048 | 0.044 | 1828 |
| 4 | 6 | 3.07 | 1007.07 | 1013.61 | 1003.92 | 0.0040 | 0.031 | 1530 |
| 5 | 9 | 0.69 | 1022.52 | 1017.45 | 1005.94 | 0.0072 | 0.160 | 2663 |
| 6 | 9 | 1.36 | 1014.35 | 1017.27 | 1006.18 | 0.0069 | 0.079 | 1873 |
| 7 | 9 | 1.42 | 1013.07 | 1017.07 | 1003.55 | 0.0084 | 0.093 | 2657 |
| 8 | 9 | 2.84 | 1007.07 | 1013.83 | 1003.40 | 0.0065 | 0.036 | 1653 |
| 9 | 9 | 2.69 | 1007.93 | 1015.73 | 1003.88 | 0.0074 | 0.040 | 1773 |
| 10 | 12 | 0.79 | 1023.05 | 1019.39 | 1005.59 | 0.0114 | 0.169 | 2517 |
| 11 | 12 | 1.25 | 1011.16 | 1013.49 | 1001.81 | 0.0097 | 0.091 | 2756 |
| 12 | 12 | 2.02 | 1009.05 | 1014.08 | 1004.11 | 0.0083 | 0.048 | 1960 |
| 13 | 12 | 3.07 | 1009.42 | 1016.12 | 1006.18 | 0.0082 | 0.031 | 1180 |
| 14 | 18 | 1.35 | 1012.00 | 1014.94 | 1003.58 | 0.0140 | 0.082 | 2442 |
| 15 | 18 | 1.91 | 1008.60 | 1014.60 | 1002.01 | 0.0155 | 0.064 | 2264 |
| 16 | 18 | 3.51 | 1007.33 | 1015.84 | 1003.95 | 0.0146 | 0.033 | 1622 |

A series of lock-exchange gravity current experiments were conducted in a rectangular plexiglass flume with linearly stratified salt water, as shown in Fig. 1. A brief introduction of the experimental set-up and procedure is summarized here. One can refer to He et al. (2017) for details. A locked head tank used to store dense salt water was placed at one side of the flume, connected by inclined perspex boards of different lengths to create different slopes. Before each experiment, linearly stratified water was gradually filled in the flume to a water depth of 34 cm using a two-tank system (Ghajar and Bang 1993). During this process, the opening height of lock 1 was kept as 4 cm and lock 2 was closed. In all the experiments, dense saline water dyed with permanganate was gently injected into the head tank to a height of 9 cm. By suddenly lifting up lock 2, the dense fluid could intrude into the ambient linearly stratified water and propagate along the slope. Lock 1 was set in front of lock 2 so it could lessen the water level fluctuation arising from the lifting of lock 2. A digital camcorder at a frame rate of 25 fps with a resolution of 4928 pixel × 3264 pixel was employed to obtain an overall view of the developmental process. A total of 16 experimental runs were conducted under different conditions as given in Table 1.

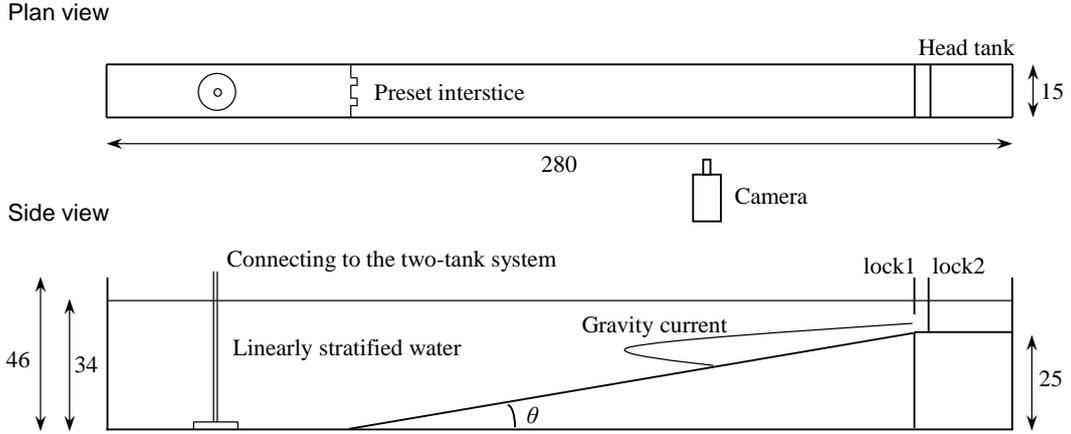

**Fig. 1**. Plan and side views of the experimental set-up (Unit: cm).

*Parameters and Front Velocity*

The ambient stratification in the flume is described by the density gradient, defined as

$$m = \frac{(\rho_B - \rho_s)\sin\theta}{\rho_s H_a}, \tag{1}$$

where $\rho_s$ is the density of the ambient fluid at the start point of the slope; $\rho_B$ is the density of the ambient fluid at the bottom of the flume; $H_a$ is the vertical distance between the gate and the bottom of the flume; and $\theta$ is the slope angle.

For a gravity current descending down a slope with limited length, the relative stratification $S$ can be used to determine whether the gravity current can separate from the slope, which is defined as (He et al. 2017)

$$S = \frac{\rho_B - \rho_s}{\rho_{c0} - \rho_s}, \tag{2}$$

where $\rho_{c0}$ is the density of the initial dense fluid. For $S > 1$, the gravity current can separate from the slope at the neutral density level where the density contrast between the current and the ambient water vanishes and then intrudes into the environment horizontally (Cortés et al. 2014; He et al. 2017).

The bulk Reynolds number Re is defined as (Dai 2013)

$$\text{Re} = \frac{\sqrt{g_0' h_l} \cdot h_l}{\nu}, \tag{3}$$

where $g_0' = g(\rho_{c0} - \rho_s) / \rho_{c0}$ is the initially reduced gravity; $h_l$ is the opening height of lock 1 and $\nu$ is the kinematic viscosity of water. In all the runs in this study, the Reynolds number is larger than 1100 (as listed in Table 1, ranging from 1180 to 3515), which means that the flow is turbulent, the viscous effect can be ignored (Dai 2013) and the motion of the current is dominated by the gravity body force.

The front location $X_f$ is defined as the distance from the start point of the slope to the forefront of the current. The corresponding front velocity $U_f$ can be found by $U_f = dX_f / dt$. Fig. 2 shows the typical time evolution of the front velocity of a gravity current in an inclined and linearly stratified environment. Once the lock is lifted, the gravity current is released and starts to accelerate along the slope. A velocity shear is then produced between the current and the ambient fluid, which generates Kelvin–Helmholtz instabilities at the upper interface (Baines 2001). The ambient lighter water is entrained into the downslope current. Meanwhile, the increase of the ambient water density quickly reduces the density contrast between the current and the ambient. These two effects lead the gravity current to accelerate in a shorter period than that in the uniform ambient. With the reduction of the density contrast, the downslope current then begins to decelerate. In addition, the front velocity also presents a fluctuation (e.g., $20 < t < 30$, $35 < t < 47$ in Fig. 2), which has been proved to be a result of the changing shape of the dense front with time and the three-dimensional action of the cross-stream water entrainment (Ieong et al. 2006). This phenomenon can also be seen in previous studies (Dai 2013; Snow and Sutherland 2014).

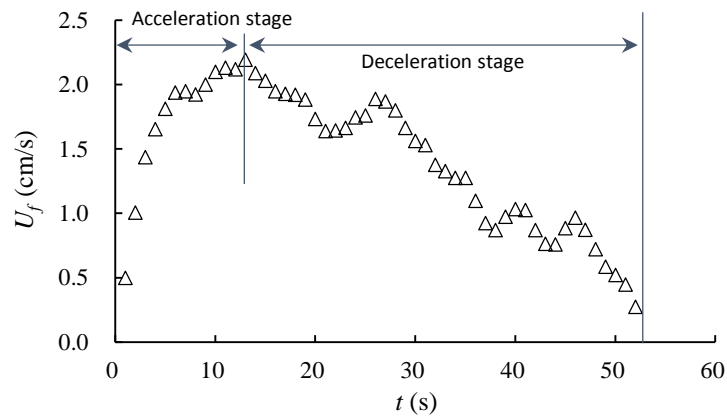

**Fig. 2.** Typical development of the front velocity ($U_f$) of a gravity current down a slope in a linearly stratified environment (Run 4).

When the density contrast vanishes, the current stops descending along the slope (Baines 2001; Guo et al. 2014). It comes to a separation stage in which the gravity current separates from the slope and then intrudes into the ambient environment horizontally at a quite low speed with a thin and sharp forward motion. For the predication of the vertical separation depth, one can refer to the work of previous researchers (Wells and Nadarajah 2009; Snow and Sutherland 2014; He et al. 2017) for more details. Although the downslope current inevitably excites internal waves, prior research has demonstrated that these internal waves have little effect on the front velocity until the current separates from the slope (Snow and Sutherland 2014). Since the current has left the slope in the separation stage, we only focus on the front velocity in the acceleration and deceleration stages and do not consider the influence of internal waves in the present study.

## Formulae of front velocity and front location

The thermal theory (Fig. 3) was first proposed by Beghin et al. (1981) to determine the development of gravity currents down a slope in a uniform ambient. In this section, we revisit this theory and extend it to describe the evolution process of lock-exchange gravity currents down a slope in linearly stratified environments.

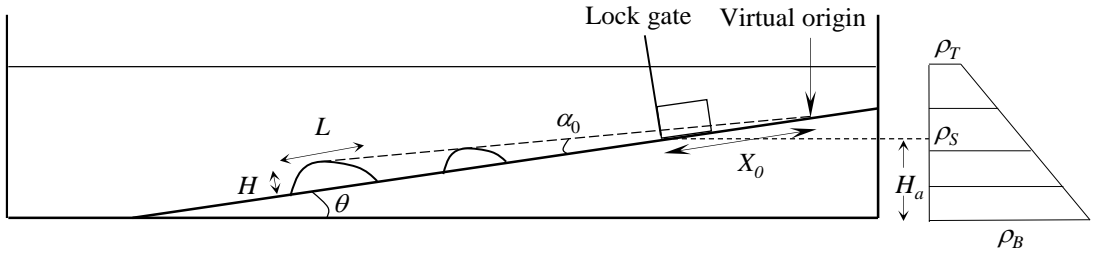

**Fig. 3.** Sketch of a gravity current in an inclined and linearly stratified environment. $\rho_T$ is the density of the ambient fluid at the top. $H$ and $L$ are the height and length of the semi-elliptical head, respectively. $X_0$ is the distance from the gate to the virtual origin. $\alpha_0$ is the growth angle of the head.

There are two main assumptions in the thermal theory. Firstly, the gravity current is assumed to be developed from a 'virtual origin' located behind the gate, which is determined by the slope and the growth of the current, as shown in Fig. 3. Secondly, the theory assumes that the head of a gravity current keeps a semi-elliptical shape, with a constant ratio of head depth $H$ to head length $L$. This study adopts these two assumptions. Therefore, the linear momentum equation of the gravity current, ignoring the bottom friction, is (Beghin et al. 1981)

$$\frac{d(\rho_a + k_v\rho_a)S_1 H L U_m}{dt} = B_c \sin\theta, \qquad (4)$$

where $t$ is time, $U_m$ is the mass-center velocity of the current head, $\rho_a$ is the density of the ambient fluid at the position of the mass-center of the current head, $k_v = 2k = 2H/L$ is the added mass coefficient (Batchelor 2000), and $S_1 = \pi/4$ is the shape factor with which the sectional area of the head is calculated by $S_1 H L$ (Dai 2013). $B_c$ is the buoyancy contained in the head of the current, which is expressed by (Dai 2013):

$$B_c = f(\rho_{c0} - \rho_a)gA_0, \tag{5}$$

where $f$ is a fraction factor and $A_0$ is the volume of the initial dense fluid which can flow down the slope.

Note that, in a linearly stratified environment in this study, $\rho_a$ should be determined by

$$\rho_a = \rho_s[1 + (X - X_0)m], \tag{6}$$

where $X$ is the distance from the virtual origin to the mass-center of the current head.

By introducing the entrainment ratio $E$, the entrainment velocity $U_e$ can be defined by (Ellison and Turner 1959)

$$U_e = EU_m. \tag{7}$$

The mass conservation equation of the gravity current has the form (Beghin et al. 1981)

$$\frac{d}{dt}(S_1 HL) = S_2(HL)^{0.5} U_e, \tag{8}$$

where $S_2 = (\pi/2^{1.5})(4k^2+1)^{0.5}/k^{0.5}$; this is another shape factor, by which the circumference of the semi-elliptical head is determined by $S_2(HL)^{0.5}$ (Dai 2013). The entrainment ratio $E$ is related with $\alpha_0$ by $E = 2\alpha_0 S_1/(k^{0.5} S_2)$ (Dai 2013).

Substituting Eq. (7) into Eq. (8) and then integrating it leads to (Beghin et al. 1981)

$$H = \frac{1}{2}\frac{S_2}{S_1}k^{0.5}EX \quad \text{and} \quad L = \frac{1}{2}\frac{S_2}{S_1}k^{-0.5}EX. \tag{9}$$

Substituting Eqs. (5), (6) and (9) into Eq. (4), the momentum equation can be written as

$$(1-X_0m)\frac{U_m d(U_m X^2)}{dX} + m\frac{U_m d(U_m X^3)}{dX} = R(\rho_{c0} - \rho_s + X_0 m) - RmX, \quad (10)$$

where $R = \dfrac{4S_1 \sin\theta f g A_0}{(1+k_v)E^2 S_2^2 \rho_s}$.

By integrating Eq.(10), one can then obtain the mass-center velocity of gravity current in the following form:

$$U_m^2 = \frac{X_0^4 U_{f0}^2}{X^4(1-mX_0+mX)^2} - \frac{2X_0^3 R(1-mX_0)(\rho_{c0}-\rho_s+\rho_s mX_0)}{3X^4(1-mX_0+mX)^2}$$
$$+ \frac{2R(1-mX_0)(\rho_{c0}-\rho_s+\rho_s mX_0)}{3X(1-mX_0+mX)^2} + \frac{mR(\rho_{c0}-\rho_s+2\rho_s mX_0)}{2(1-mX_0+mX)^2}, \quad (11)$$
$$- \frac{X_0^4 mR(\rho_{c0}-\rho_s+2\rho_s mX_0)}{2X^4(1-mX_0+mX)}$$

where $U_{f0}$ is the initial mass-center velocity of the current. As the front velocity $U_f$ is much easier to be measured than the mass-center velocity $U_m$, by substituting $U_f = (1 + \alpha_0/2k)U_m$ (Dai 2013) into Eq. (11), the relationship between the front velocity $U_f$ and front location of lock-exchange gravity currents down a slope in linearly stratified environments can be easily obtained:

$$U_f^2 = (1+\alpha_0/2k)^2 U_m^2 = (1+\alpha_0/2k)^2 [\frac{X_0^4 U_{f0}^2}{X^4(1-mX_0+mX)^2}$$
$$-\frac{2X_0^3 R(1-mX_0)(\rho_{c0}-\rho_s+\rho_s mX_0)}{3X^4(1-mX_0+mX)^2} + \frac{2R(1-mX_0)(\rho_{c0}-\rho_s+\rho_s mX_0)}{3X(1-mX_0+mX)^2} . \quad (12)$$
$$+ \frac{mR(\rho_{c0}-\rho_s+2\rho_s mX_0)}{2(1-mX_0+mX)^2} - \frac{X_0^4 mR(\rho_{c0}-\rho_s+2\rho_s mX_0)}{2X^4(1-mX_0+mX)}]$$

Note that, when the ambient environment is uniform, i.e. the ambient stratification $m = 0$, Eq. (11) can be simplified as

$$U_m^2 = U_{f0}(\frac{X_0}{X})^4 + \frac{8S_1 \sin\theta B_c}{3X(1+k_v)E^2 S_2^2 \rho_s}[1-(\frac{X_0}{X})^3]. \quad (13)$$

Eq. (13) is the same as the result from Beghin et al. (1981). If the gravity current starts from a quiescent state, the initial mass-center velocity $U_{f0}$ can be assumed to be zero (Dai 2013). One can get the front velocity of lock-exchange gravity currents down a slope in unstratified environments as:

$$U_f = (1+\frac{\alpha_0}{2k})\sqrt{[1-(\frac{X_0}{X})^3]\frac{1}{3X}}\sqrt{\frac{8\sin\theta S_1(\rho_{c0}-\rho_a)gfA_0}{(1+k_v)E^2 S_2^2 \rho_a}} . \qquad (14)$$

In Eq. (12), there are several parameters that need to be determined to predict the relationship between the front velocity and front location of lock-exchange gravity currents down a slope in linearly stratified environments. The parameters $m$, $\rho_{c0}$, $\rho_s$, $S_1$, and $\theta$, are related to the initial experimental conditions and can be directly measured. The parameters $\alpha_0$, $k = H/L$, $X_0$, $U_{f0}$ can be measured during the motion of the gravity current. The parameters $k_v = 2k$ and $E = 2\alpha_0 S_1/(k^{0.5} S_2)$ are calculated using the values of the above parameters. However, the parameter $f$, defined as the fraction of the heavy fluid contained in the head of the gravity current, is difficult to be determined. This fraction is important because it is used to calculate the buoyancy ($B_c$) contained in the head of the current. Its value varies during the propagation of a gravity current in stratified water. First, this fraction changes due to entrainment and mixing with ambient water. Second, the fraction in the deceleration stage should be much smaller than that in the acceleration stage since the dense fluids in the downslope current at different depths attempt to find their own neutral density levels and detrain into the environment during the deceleration stage (He et al. 2017). Because $f$ is different in the accelerating and the decelerating stage, we simplify Eq. (12) in the following in order to obtain easy-to-use equations for each regime that may find application in future studies.

*Front velocity of gravity current in a linearly stratified ambient in the acceleration stage*

As the mechanisms and the dynamic features of the gravity current in the two stages are greatly different, following the previous researchers (Beghin et al. 1981; Dai 2013), we apply different methods in the acceleration and deceleration stages to further simplify the formula. In the acceleration stage, the lock-exchange gravity current in an inclined and linearly stratified environment only moves a short distance along the depth. Meanwhile, the gravity current is not fully developed so the length of its head is relatively small. Therefore, for the acceleration stage, the following simplifications can be further applied:

(1) The gravity current starts from a quiescent state, so $U_{f0}=0$ (Dai 2013);

(2) The order of magnitude of $mX_0$, i.e., $\dfrac{(\rho_B-\rho_s)}{\rho_s}\dfrac{X_0\sin\theta}{H_a}$, is much smaller than that of $\rho_{c0}-\rho_s$, so $\rho_{c0}-\rho_s+mX_0\approx\rho_{c0}-\rho_s$;

(3) During the acceleration stage, the length of head of the current is relatively small, consequently, $X_f\approx X-X_0$;

(4) Compared to the total vertical distance $H_a$, the vertical movement distance of the current $X_f\sin\theta$ in the acceleration stage is relatively small, so $\dfrac{X_f\sin\theta}{H_a}\ll 1$.

By applying the above simplifications (1-4) and the relationship of $1-mX_0+mX=\dfrac{\rho_a}{\rho_s}\approx 1+\dfrac{(\rho_B-\rho_s)}{\rho_s}\dfrac{X_f\sin\theta}{H_a}$ in Eq.(12), one can get the front velocity of a gravity current in linearly stratified ambience as:

$$U_f^2=(1+\alpha_0/2k)^2\{2R(1-mX_0)(\rho_{c0}-\rho_s)\dfrac{X_f}{(X_f+X_0)^2}\\+\dfrac{mR(\rho_{c0}-\rho_s)}{2}[1-(\dfrac{X_0}{X_f+X_0})^4]\}. \quad (15)$$

In this step, the expression associated with the location of the current $\left[1-\left(X_0/(X_f+X_0)\right)^3\right]/3X$ is approximated by $X_f/(X_f+X_0)^2$ according to the suggestion of Beghin et al. (1981). Similarly, we can approximate $\left[1-\left(X_0/(X_f+X_0)\right)^4\right]/2$ with $2X_f^2/(X_f+X_0)^2$ to further simplify Eq. (15) as:

$$U_f^2 \approx (1+\frac{\alpha_0}{2k})^2 \frac{X_f}{(X_f+X_0)^2} 2R(\rho_{c0}-\rho_S)\alpha_s, \tag{16}$$

where the new parameter, i.e., the stratification coefficient $\alpha_s$ is expressed by:

$$\alpha_s = 1 - mX_f \frac{\rho_S}{\rho_{c0}-\rho_S} = \frac{g'}{g_0'}, \tag{17}$$

where $g' = g(\rho_{c0} - \rho_a) / \rho_{c0}$ is the reduced gravitational acceleration at the head of the gravity current. Note that $\alpha_s$ considers the influence of the ambient stratification on the movement of the gravity current at the acceleration stage.

The fraction factor $f$ in Eq. (16) was difficult to determine, and was assumed to be unity by Beghin et al. (1981), or estimated by fitting with the experimental data by Dai (2013). In all previous experiments related to the thermal theory (Beghin et al. 1981; Rastello and Hopfinger 2004; Maxworthy and Nokes 2007; Maxworthy 2010; Dai 2013), the experimental tanks were almost the same, in that locks were set vertically to the slope (see Fig. 3). However, as the present experiments were conducted in a linearly stratified environment, a similar set-up to that in Baines (2001) was adopted, in which the head tank was set horizontally (see Fig. 1). Under this set-up, the experimental results show that only a small part of the initial dense fluids in the horizontal tank could flow down the slope. These different experimental conditions and influential mechanisms make the values of $f$ and $A_0$ different from those in the previous studies. It is hard to determine these two values at the same time. To avoid this problem, a new parameter, the geometric configuration coefficient $c_a$, is introduced. We can rewrite Eq. (16):

$$U_f = \frac{dX_f}{dt} = \frac{P\sqrt{X_f(1-WX_f)}}{X_f + X_0}, \qquad (18)$$

where

$$P = (1+\frac{\alpha_0}{2k})\sqrt{\frac{8\sin\theta S_1(\rho_{c0}-\rho_s)g(c_a A_0)}{(1+k_v)E^2 S_2^2 \rho_s}} \quad \text{and} \quad W = \frac{m\rho_s}{\rho_{c0}-\rho_s}. \qquad (19)$$

In this way, $f$ and $A_0$ can also be assumed to be unity and the total volume of the initial dense fluid respectively, as done in the previous study (Beghin et al. 1981). As $f$ is set as unity, we do not put it into the equations anymore. The geometric configuration coefficient $c_a$ then can be treated as the combined effect of the experimental configuration and different mechanisms on $f$ and $A_0$. Eqs. (18) and (19) and are the formulae for the front velocity and front location of lock-exchange gravity currents in an inclined and linearly stratified environment in the acceleration stage.

As the dominant driving force controlling the motion and the dynamic features of the current are different in the acceleration and deceleration stages, it is necessary to determine where the transition point between the two stages is. Theoretically, assuming that the current turns to the deceleration stage right after reaching the maximum velocity in the end of the acceleration stage, the transition point (i.e., $X_{f,p}$) is defined as the linkage between the acceleration and deceleration stage. $X_{f,p}$ can be determined by setting the derivative with respect to distance of Eq. (18) equal to be zero. Thus, the transition point $X_{f,p}$ is calculated as

$$X_{f,p} = \frac{X_0}{1+2X_0 W}. \qquad (20)$$

The maximum front velocity then can be determined by taking Eq. (20) into Eq. (18):

$$U_{f,\max} = \frac{P\sqrt{X_0(1+WX_0)}}{X_0 + X_0(1+2WX_0)}. \tag{21}$$

*Front velocity of gravity current in a linearly stratified ambient in the deceleration stage*

When the gravity current is sufficiently far into the deceleration stage, its propagation distance becomes longer. In addition to the simplifications (1) and (2), the following simplification is adopted (Dai 2013):

$$\frac{X}{X_0} \gg 1. \tag{22}$$

Meanwhile, according to the definition, front location $X_f$ and $X$ has the relationship as (Dai 2013)

$$X_f + X_0 = (1 + \frac{\alpha_0}{2k})X. \tag{23}$$

By substituting Eqs. (22) and (23) into Eq. (12), He et al. (2017) has derived the front velocity in the deceleration stage. To keep consistency, this paper summarizes it as

$$U_f = \frac{dX_f}{dt} = \frac{\sqrt{I/(X_f + X_0) + J}}{M + G(X_f + X_0)}, \tag{24}$$

where

$$I = \frac{8S_1 \sin\theta g(c_d A_0)}{3(1+k_v)E^2 S_2 \rho_S}(1+\frac{\alpha_0}{2k})(\rho_{c0} - \rho_S - \rho_S m^2 X_0^2), \quad G = \frac{m}{(1+\alpha_0/2k)^2}, \tag{25}$$

and

$$J = \frac{2mS_1 \sin\theta g(c_d A_0)}{(1+k_v)E^2 S_2 \rho_S}(\rho_{c0} - \rho_S + 2mX_0), \quad M = \frac{1}{1+\alpha_0/2k}. \tag{26}$$

Similarly, the geometric configuration coefficient $c_d$ in the deceleration stage has been considered in $I$ and $J$.

Eqs. (24), (25) and (26) are the formulae to determine the front velocity and front location of the lock-exchange gravity currents down a slope in linearly stratified environments in the deceleration stage.

Therefore, Eqs. (24) - (26) together with Eqs. (18) and (19) form the complete formulae to describe the whole propagation of lock-exchange gravity currents along a slope in a linear stratification. Eqs. (20) and (21) are used to determine the transition point between the acceleration and deceleration stages and the corresponding maximum front velocity.

## Validation of the proposed formulae

The measured data of $X_f$ and $U_f$ from the present experiment are employed to validate the above formulae. For each experiment, we first determine $c_a$ and $c_d$ by setting the best-fitting lines through plots of the measured front velocity versus front location in the respective acceleration and deceleration stages. Then, the equations of the relationship between the front location and time are solved and the results are compared with the experimental data. The consistency check towards Eq. (12) is also performed by comparing the calculated front velocity, in which the fitted $c_a$ and $c_d$ from the first step are used, with the experimental data.

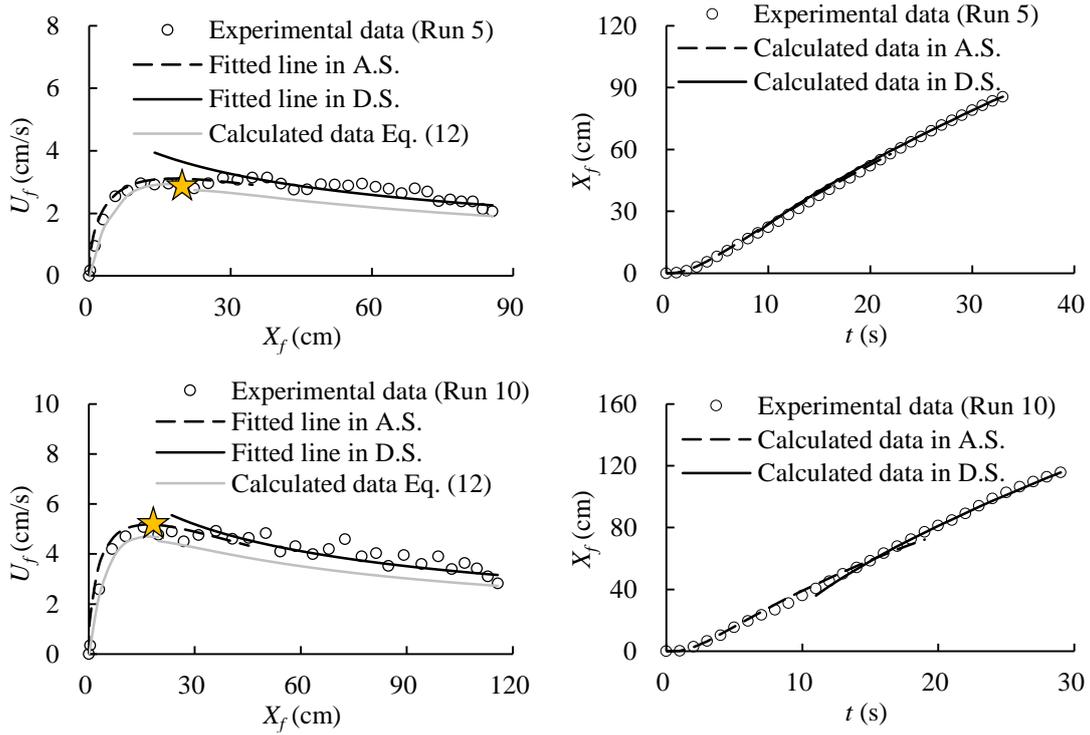

**Fig. 4.** Validation of the simplified formulae and consistency check of the unified formula. In weak stratification. A.S. and D.S. mean the acceleration stage and the deceleration stage, respectively. ☆ indicates the maximum front velocity and the corresponding turning point.

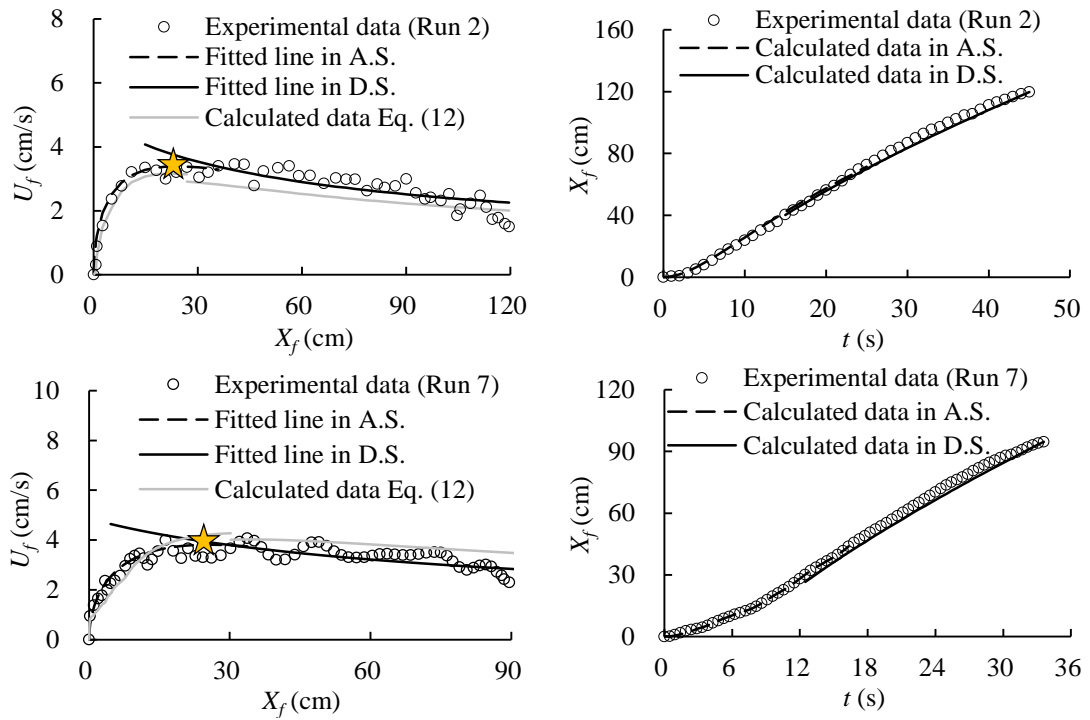

**Fig. 5.** Validation of the simplified formulae and consistency check of the unified formula. In weak-medium stratification.

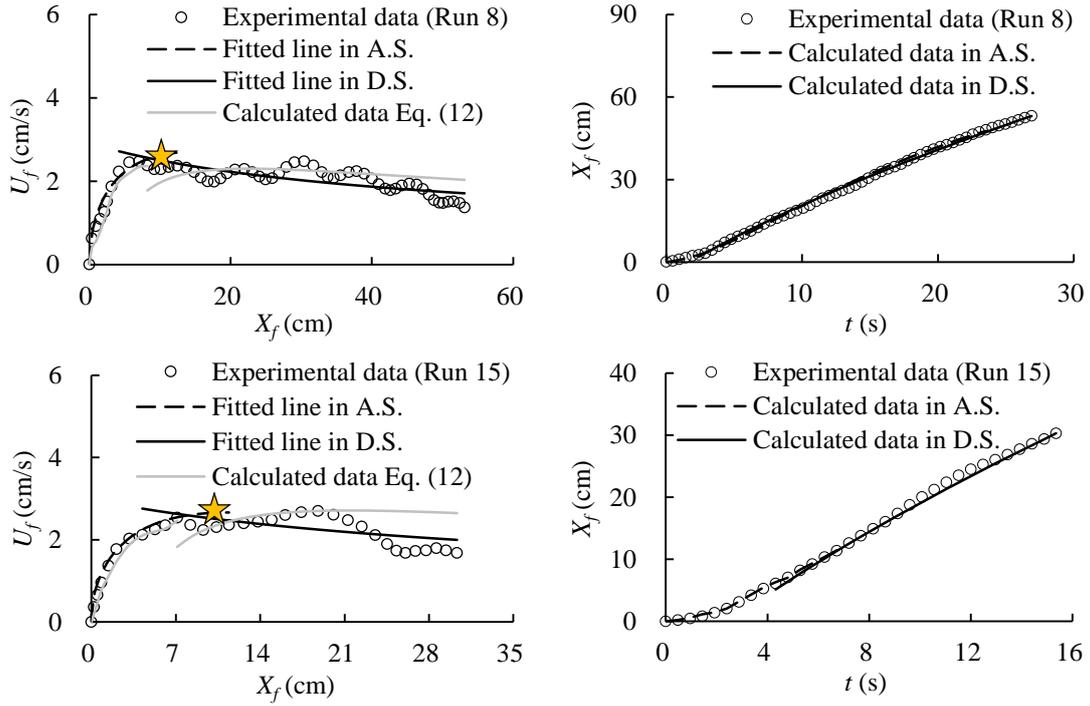

**Fig. 6.** Validation of the simplified formulae and consistency check of the unified formula. In medium-strong stratification.

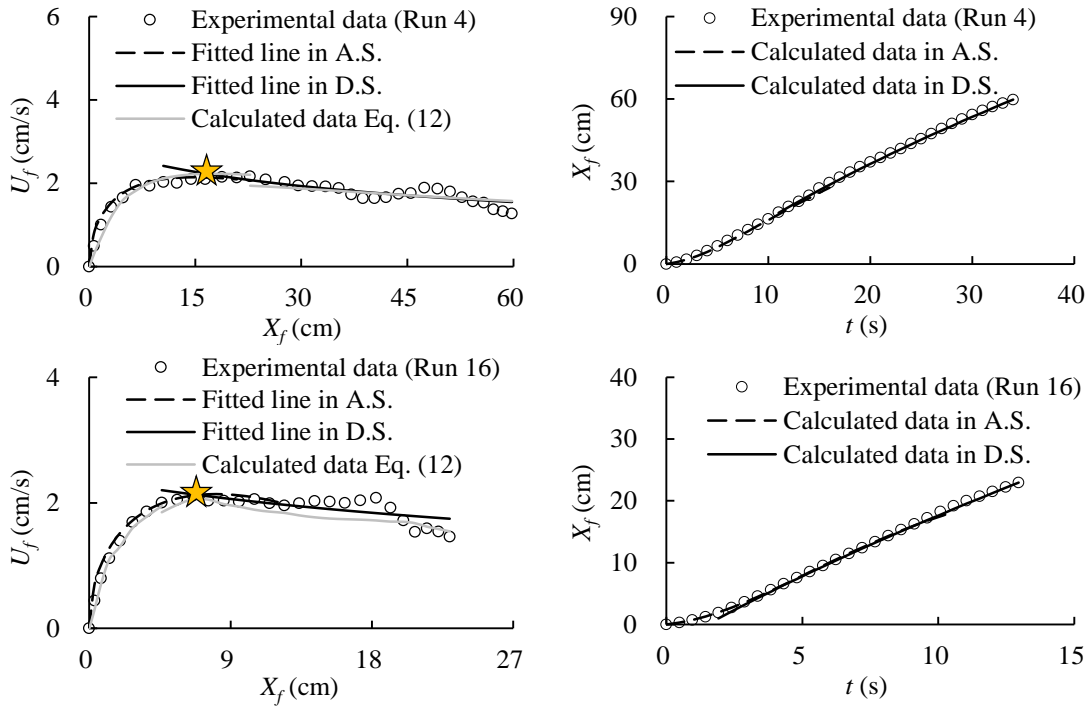

**Fig. 7.** Validation of the simplified formulae and consistency check of the unified formula. In strong stratification.

The parameters in all 16 experimental runs are listed in Table 2. Eight comparisons between the experimental data and calculated data are shown in Figs. 4-7 for convenience, in which, Fig. 4 represents the cases with weak stratification, Fig. 5 shows the cases with medium-weak stratification, Fig. 6 presents the cases with medium-strong stratification, and Fig. 7 shows the cases with strong stratification. It can be seen that the acceleration and deceleration propagations and the maximum front velocity with the transition point can be well described by the proposed formulae. The unified equation, i.e., Eq. (12), is also validated. From the calculation, we also notice the value of $J$ is much smaller than the other term $I/(X_f + X_0)$, as shown in Table 2, so Eq. (24) can be further simplified into

$$U_f = \frac{dX_f}{dt} = \frac{\sqrt{I/(X_f + X_0)}}{M + G(X_f + X_0)} \tag{27}$$

**Table 2**. The relevant parameters in the proposed formulae

| Run | $a_0$ | $k$ | $E$ | $X_0$ (m) | $c_d$ | $c_a$ | $I$ (m³/s²) | $G$ (1/m) | $M$ | $J$ (m²/s²) | $P$ (m^1.5/s) | $W$ (1/m) |
|---|---|---|---|---|---|---|---|---|---|---|---|---|
| 1 | 0.044 | 0.28 | 0.054 | 0.30 | 0.022 | 0.035 | 0.0017 | 0.0038 | 0.93 | 0.0000048 | 0.060 | 0.27 |
| 2 | 0.051 | 0.37 | 0.059 | 0.32 | 0.025 | 0.045 | 0.0007 | 0.0034 | 0.93 | 0.0000016 | 0.042 | 0.55 |
| 3 | 0.053 | 0.28 | 0.064 | 0.35 | 0.017 | 0.045 | 0.0003 | 0.0040 | 0.92 | 0.0000007 | 0.031 | 1.07 |
| 4 | 0.050 | 0.28 | 0.064 | 0.24 | 0.016 | 0.040 | 0.0002 | 0.0034 | 0.92 | 0.0000003 | 0.024 | 1.29 |
| 5 | 0.052 | 0.32 | 0.104 | 0.21 | 0.024 | 0.039 | 0.0023 | 0.0061 | 0.92 | 0.0000011 | 0.030 | 0.43 |
| 6 | 0.044 | 0.33 | 0.052 | 0.30 | 0.010 | 0.022 | 0.0006 | 0.0061 | 0.94 | 0.0000026 | 0.043 | 0.85 |
| 7 | 0.059 | 0.31 | 0.071 | 0.47 | 0.024 | 0.071 | 0.0009 | 0.0070 | 0.91 | 0.0000048 | 0.062 | 0.89 |
| 8 | 0.066 | 0.24 | 0.084 | 0.28 | 0.014 | 0.065 | 0.0002 | 0.0050 | 0.88 | 0.0000006 | 0.034 | 1.52 |
| 9 | 0.060 | 0.40 | 0.066 | 0.41 | 0.009 | 0.040 | 0.0002 | 0.0064 | 0.93 | 0.0000006 | 0.032 | 1.83 |
| 10 | 0.079 | 0.29 | 0.096 | 0.22 | 0.021 | 0.034 | 0.0011 | 0.0089 | 0.88 | 0.0000081 | 0.051 | 0.66 |
| 11 | 0.096 | 0.40 | 0.106 | 0.25 | 0.021 | 0.051 | 0.0004 | 0.0077 | 0.89 | 0.0000026 | 0.041 | 1.04 |
| 12 | 0.070 | 0.26 | 0.088 | 0.37 | 0.012 | 0.054 | 0.0002 | 0.0064 | 0.88 | 0.0000010 | 0.037 | 1.68 |
| 13 | 0.084 | 0.37 | 0.095 | 0.25 | 0.015 | 0.051 | 0.0001 | 0.0066 | 0.90 | 0.0000006 | 0.027 | 2.55 |
| 14 | 0.113 | 0.35 | 0.132 | 0.26 | 0.016 | 0.057 | 0.0003 | 0.0104 | 0.86 | 0.0000026 | 0.042 | 1.67 |
| 15 | 0.113 | 0.32 | 0.135 | 0.25 | 0.010 | 0.046 | 0.0002 | 0.0112 | 0.85 | 0.0000013 | 0.033 | 2.36 |
| 16 | 0.131 | 0.45 | 0.137 | 0.27 | 0.017 | 0.105 | 0.0001 | 0.0112 | 0.87 | 0.0000008 | 0.033 | 4.35 |

**Discussion**

The propagation of lock-exchange gravity currents down a slope in linearly stratified environments is very complicated due to entrainment and mixing between the current and the stratified ambient. During the initial time (acceleration stage), the density difference between the heavy current and the light ambient water is the main factor driving its movement. As the current moves further down the slope, the density difference gradually decreases due to the entrainment effect and the density increase of the ambient water. In the present formulae, these two factors are well reflected by the entrainment ratio $E$ and the density gradient $m$, respectively. The motion of the current is generally controlled by the buoyancy force and then viscous force. In different stages, the distance that the current propagates is greatly determined by the density contrast. Although the whole propagating process of gravity current can be described by the unified expression of Eq. (12), this formula involves several parameters that are difficult to be determined. For instance, the mixing and entrainment between the current and ambient water changes the fraction of the heavy fluid contained in the head in the acceleration and deceleration stages. Therefore, several simplifications are applied in two stages to simplify the formula to easily calculate the front velocity and front location of the current.

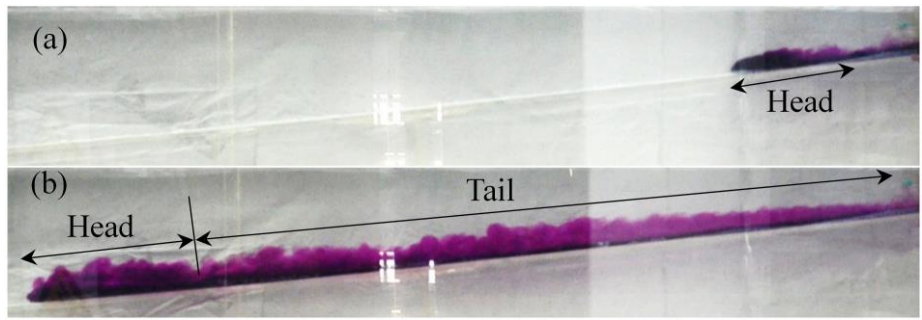

**Fig. 8.** Gravity current in an inclined and linearly stratified environment (Run 2). (a) acceleration stage, $t = 3$ s, more dense fluids are contained in the head; (b) deceleration stage, $t = 32$ s, the tail of the gravity current gets thick and more dense fluids stay in the tail.

In fact, the value of parameter $f$ (i.e. the fraction of the heavy fluid contained in the head) has varied greatly in previous studies of gravity currents in unstratified environments. It was assumed to be unity by Beghin et al. (1981), while it was fitted to be about 0.4 by Maxworthy (2010) and about 0.8 by Dai (2013). In a stratified environment, the situation is much more complicated as the evolution process is significantly influenced not only by the entrainment but also by the vertical stratification. By introducing geometric configuration coefficients $c_a$ and $c_d$ in the present formulae, we do not directly determine the value of *f* but assume it to be unity as done by Beghin et al. (1981). The data in Table 2 show that $c_a$ is larger than $c_d$ in all the experimental cases, which implies that the actual fraction of the heavy fluid contained in the head in the acceleration stage is also larger than that in the deceleration stage. This is because the dominant driving force of gravity currents in the acceleration and deceleration stages is different. In the acceleration stage, the density contrast between the gravity current and the ambient water is sufficiently large to drive the current down the slope so the head contains a larger fraction of the buoyancy (see Fig. 8a). Subsequently, the density of the downslope current decreases due to entrainment and density increase of the ambient water. Furthermore, the dense fluids in the downslope current at different depths attempt to find their own neutral density levels and detrain into the environment (He et al. 2017). Thus, a large fraction of the dense fluids stays within the tail (see Fig. 8b). Consequently, the head of the gravity current contains a smaller fraction of dense fluid in the deceleration stage.

One of the main assumptions in the thermal theory is that the gravity current was developed from a 'virtual origin', which is determined by the slope and the growth of the head height. When the gravity current propagates on a horizontal plane, the height of the head does not increase with distance so the assumption of the virtual origin might not be applied (Dai 2013). The previous researchers (Beghin et al. 1981; Dai 2013) have indicated that the thermal theory is not applicable for gravity currents on a horizontal boundary. Similarly, the present formulas have the same limitation. For the theory in this situation, the reader can refer to the work of Maxworthy et al. (2002)

and Ungarish (2006). However, the assumptions and derivations are not limited to larger slope angles so the results in the present study are essentially suitable for steep slopes, though the specific parameters may have to be re-calibrated. The equations developed in this study have been validated using experimental data with a slope varied from 6° to 18°. It is suggested that further experiments on more steep slopes should be conducted to investigate the lock-exchange gravity currents in linearly stratified environments.

## Conclusion

This study presents a complete set of analytical formulae to determine the front velocity and front location of lock-exchange gravity currents down a slope in linearly stratified environments. The formulae are developed from mass conversation and momentum equations based on the thermal theory, by further considering the vertical linear stratification of ambient water, i.e. parameter $m$. The lock-exchange experiments show the evolution of the gravity current can be distinguished as a short acceleration stage and then a deceleration stage based on its front velocity before it leaves the slope. In the acceleration stage, the formula for front velocity takes into account the influence of the ambient stratification by the stratification coefficient $\alpha_s$. As for the deceleration stage, the $U_f$ - $X_f$ relationship is derived by adding a parameter which describes the density distribution of ambient water. Two geometric configuration coefficients are introduced in the formulae in the respective acceleration and deceleration stages to consider the influences of the experimental configuration and mechanisms on the fraction of the buoyancy contained in the head and the volume of the downslope current. The transition point between the acceleration and deceleration stages and the corresponding maximum front velocity can be also determined by the proposed formulae. The good agreements between the data from the experiments and the formulae validate the capacity of the proposed formulae to describe the evolution process of lock-exchange gravity currents down a slope in linearly stratified environments.

Furthermore, the formulae could also be applied to describe the development of lock-exchange gravity currents down a slope in other kinds of stratified environments by modifying the stratification parameter *m*. The present study mainly focuses on the development of particle-free gravity current down a slope in linearly stratified environments. The applicability of the present theory to particulate gravity currents needs further experimental and theoretical work in the future.

## Acknowledgement

This work was partially supported by the National Key Research and Development Program of China (2017YFC0405502), National Natural Science Foundation of China (11672267), Natural Science Foundation of Zhejiang Province (LR16E090001), and Research Funding of Shenzhen City (JCYJ20160425164642646).

## References

Baines, P. G. (2001). "Mixing in flows down gentle slopes into stratified environments." *J. Fluid Mech.*, 443, 237-270.

Batchelor, G. K. (2000). *An introduction to fluid dynamics*, Cambridge university press, Cambridge.

Beghin, P., Hopfinger, E. J., and Britter, R. E. (1981). "Gravitational convection from instantaneous sources on inclined boundaries." *J. Fluid Mech.*, 107, 407-422.

Benjamin, T. B. (1968). "Gravity currents and related phenomena." *J. Fluid Mech.*, 31(02), 209-248.

Birman, V. K., Meiburg, E., and Ungarish, M. (2007). "On gravity currents in stratified ambients." *Phys. of Fluids*, 19(8), 86602.

Cortés, A., Rueda, F. J., and Wells, M. G. (2014). "Experimental observations of the splitting of a gravity current at a density step in a stratified water body." *J. Geophys. Res. Oceans*, 119(2), 1038-1053.

Dade, W. B., Lister, J. R., and Huppert, H. E. (1994). "Fine-sediment deposition from gravity surges on uniform slopes." *J. Sediment Res.*, 64(3).

Dai, A. (2013). "Experiments on gravity currents propagating on different bottom slopes." *J. Fluid Mech.*, 731, 117-141.

Dai, A. (2014). "Non-Boussinesq gravity currents propagating on different bottom slopes." *J. Fluid Mech.*, 741, 658-680.

Dai, A. (2015). "Thermal theory for non-Boussinesq gravity currents propagating on inclined boundaries." *J. Hydraul. Eng.*, 10.1061/(ASCE)HY.1943-7900.0000949: 141(1), 6014021.

Ellison, T. H., and Turner, J. S. (1959). "Turbulent entrainment in stratified flows." *J. Fluid Mech.*, 6(3), 423-448.

Fernandez, R. L., and Imberger, J. (2008). "Time-varying underflow into a continuous stratification with bottom slope." *J. Hydraul Eng.*, 10.1061/(ASCE)0733-9429(2008)134:9(1191): 134(9), 1191-1198.

Ghajar, A. J., and Bang, K. (1993). "Experimental and analytical studies of different methods for producing stratified flows." *Energy*, 18(4), 323-334.

Guo, Y., Zhang, Z., and Shi, B. (2014). "Numerical Simulation of Gravity Current Descending a Slope into a Linearly Stratified Environment." *J. Hydraul Eng.*, 10.1061/(ASCE)HY.1943-7900.0000936: 140(12), 4014061.

He, Z., Zhao, L., Lin, T., Hu, P., Lv, Y., Ho, H., and Lin, Y. (2017). "Hydrodynamics of gravity currents down a ramp in linearly stratified environments." *J. Hydraul Eng.*, 10.1061/(ASCE)HY.1943-7900.0001242: 143(3), 4016085.

Ho, H., and Lin, Y. (2015). "Gravity currents over a rigid and emergent vegetated slope." *Adv. Water Resour.*, 76, 72-80.

Huppert, H. E., and Simpson, J. E. (1980). "The slumping of gravity currents." *J. Fluid Mech.*, 99(04), 785-799.

Ieong, K. K., Mok, K. M., and Yeh, H. (2006). "Fluctuation of the front propagation speed of developed gravity current." *J. Hydrodyn., Ser. B*, 18(3, Supplement), 351-355.

Longo, S., Ungarish, M., Di Federico, V., Chiapponi, L., and Addona, F. (2016). "Gravity currents in a linearly stratified ambient fluid created by lock release and influx in semi-circular and rectangular channels." *Phys. Fluids*, 28(9), 96602.

Maxworthy, T. (2010). "Experiments on gravity currents propagating down slopes. Part 2. The evolution of a fixed volume of fluid released from closed locks into a long, open channel." *J. Fluid Mech.*, 647, 27-51.

Maxworthy, T., Leilich, J., Simpson, J. E., and Meiburg, E. H. (2002). "The propagation of a gravity current into a linearly stratified fluid." *J. Fluid Mech.*, 453, 371-394.

Maxworthy, T., and Nokes, R. I. (2007). "Experiments on gravity currents propagating down slopes. Part 1. The release of a fixed volume of heavy fluid from an enclosed lock into an open channel." *J. Fluid Mech.*, 584, 433-453.

Meiburg, E., Radhakrishnan, S., and Nasr-Azadani, M. (2015). "Modeling Gravity and Turbidity Currents: Computational Approaches and Challenges." *Appl. Mech. Rev.*, 67(4), 40802.

Ottolenghi, L., Adduce, C., Inghilesi, R., Armenio, V., and Roman, F. (2016). "Entrainment and mixing in unsteady gravity currents." *J. Hydraul Res.*, 54(5), 1-17.

Rastello, M., and Hopfinger, E. J. (2004). "Sediment-entraining suspension clouds: a model of powder-snow avalanches." *J. Fluid Mech.*, 509, 181-206.

Samothrakis, P., and Cotel, A. J. (2006). "Finite volume gravity currents impinging on a stratified interface." *Exp. Fluids*, 41(6), 991-1003.

Simpson, J. E. (1982). "Gravity currents in the laboratory, atmosphere, and ocean." *Annu. Rev. Fluid Mech.*, 14(1), 213-234.

Snow, K., and Sutherland, B. R. (2014). "Particle-laden flow down a slope in uniform stratification." *J. Fluid Mech.*, 755, 251-273.

Steenhauer, K., Tokyay, T., and Constantinescu, G. (2017). "Dynamics and structure

of planar gravity currents propagating down an inclined surface." *Phys. Fluids*, 29(3), 36604.

Ungarish, M. (2006). "On gravity currents in a linearly stratified ambient: a generalization of Benjamin's steady-state propagation results." *J. Fluid Mech.*, 548, 49-68.

Wells, M., and Nadarajah, P. (2009). "The Intrusion Depth of Density Currents Flowing into Stratified Water Bodies." *J Phys. Oceanogr.*, 39(8), 1935-1947.

28